\documentclass[runningheads]{llncs}
\usepackage{graphicx}
\usepackage{amsmath}
\usepackage{amssymb}
\usepackage{dblfloatfix}
\usepackage[colorlinks]{hyperref}
\usepackage{tikz}
\usepackage{tabularx}

\usepackage{xcolor}

\begin{document}


%
\title{3D Kidneys and Kidney Tumor Semantic Segmentation using Boundary-Aware Networks}
\titlerunning{3D Kidney Tumor Segmentation using Boundary-Aware Networks}
\author{
Andriy Myronenko\inst{1} \and
Ali Hatamizadeh\inst{1,2}
}

\authorrunning{A.~Hatamizadeh et al.}
\authorrunning{A.~Myronenko, and A.~Hatamizadeh}
\institute{NVIDIA, Santa Clara, CA, USA \and Computer Science Department, University of California, Los Angeles, CA, USA  \\ \email{\{amyronenko,ahatamizadeh\}@nvidia.com} }
\maketitle 

\begin{abstract}
Automated segmentation of kidneys and kidney tumors is an important step in quantifying the tumor's morphometrical details to monitor the progression of the disease and accurately compare decisions regarding the kidney tumor treatment. Manual delineation techniques are often tedious, error-prone and require expert knowledge for creating unambiguous representation of kidneys and kidney tumors segmentation. In this work, we propose an end-to-end boundary aware fully Convolutional Neural Networks (CNNs) for reliable kidney and kidney tumor semantic segmentation from arterial phase abdominal 3D CT scans. We propose a segmentation network consisting of an encoder-decoder architecture that specifically accounts for organ and tumor edge information by devising a dedicated boundary branch supervised by edge-aware loss terms. We have evaluated our model on 2019 MICCAI KiTS Kidney Tumor Segmentation Challenge dataset and our method has achieved dice scores of $0.9742$ and $0.8103$ for kidney and tumor repetitively and an overall composite dice score of $0.8923$.


\end{abstract}

\section{Introduction}
\label{sec:intro}

Kidney cancer accounted for nearly 175,000 deaths worldwide in 2018 \cite{bray2018global}, and it is projected that 14,770 deaths will occur due to the disease in 2019 in the US~\cite{siegel2019cancer}. Current kidney tumor treatment planning include Radical Nephrectomy (RN) and Partial Nephrectomy (PN). In RN, both the tumor and the affected kidney are removed whereas in PN the tumor is removed but kidneys are saved \cite{sun2012treatment}. Although RNs were historically prevalent as a standard treatment procedure for kidney tumors, new capabilities for earlier detection of the tumors as well as advancements in surgery has made PNs a viable treatment approach~\cite{heller2019kits19}. 

Automated segmentation of kidneys and kidney tumors assists physicians to obtain accurate morphometrical details of the tumor in an efficient and reliable manner as the manual delineation process is often tedious and error-prone. The decision for kidney tumor treatment plan can be made by leveraging such important tumor's morphometrical information. Recently, deep learning approaches for semantic image segmentation have demonstrated prominent results in medical image analysis for various applications~\cite{Milletari16,Myronenko18,hatamizadeh2019deep,hatamizadeh2019deeplesion}. The powerful non-linear feature extraction capabilities of CNNs along with the effectiveness of the encoder-decoder architectures have made it possible to employ CNNs for challenging segmentation tasks. 

In this paper, we propose a boundary-aware fully Convolutional Neural Networks for end-to-end and reliable semantic segmentation of kidneys and kidney tumor by encoding the information of edges in a dedicated stream that is supervised by edge-aware losses. We have trained and tested our model on 2019 KiTS Kidney Tumor Segmentation Challenge and results demonstrate the effectiveness of our proposed framework.   

\begin{figure*}[t!]

\includegraphics[width=0.48\linewidth,height=0.48\linewidth]{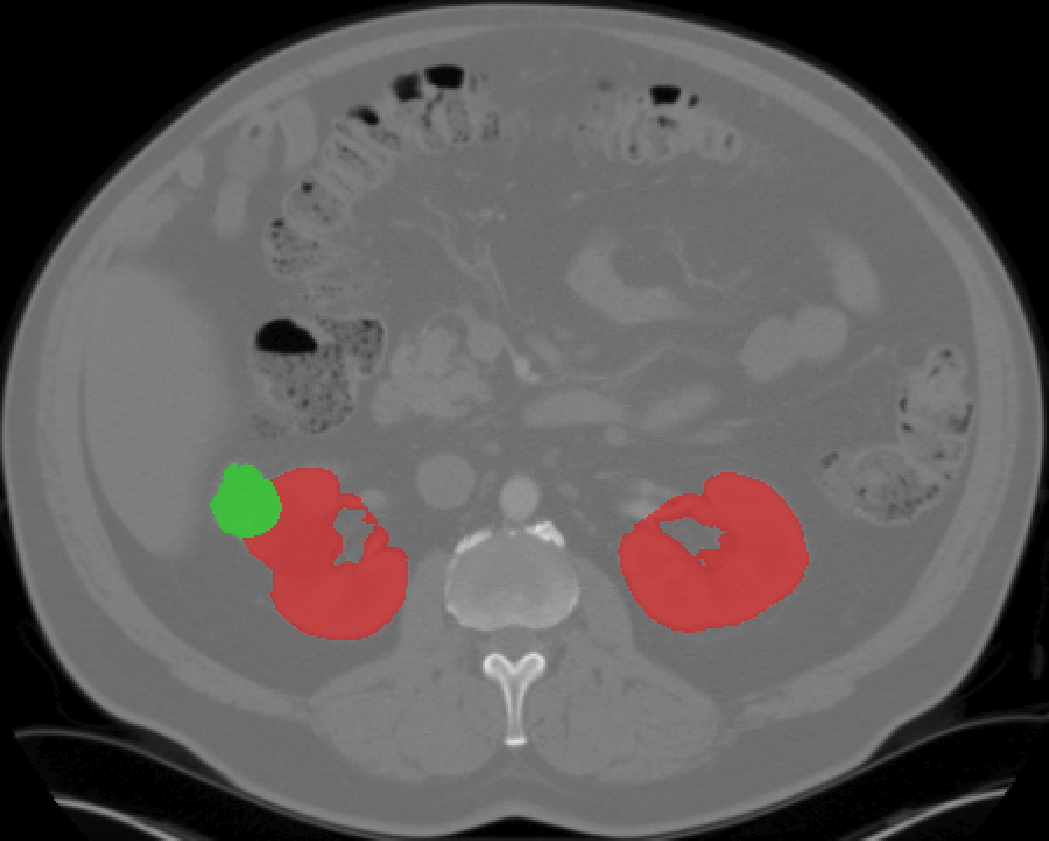}
\hfill
\includegraphics[width=0.48\linewidth,height=0.48\linewidth]{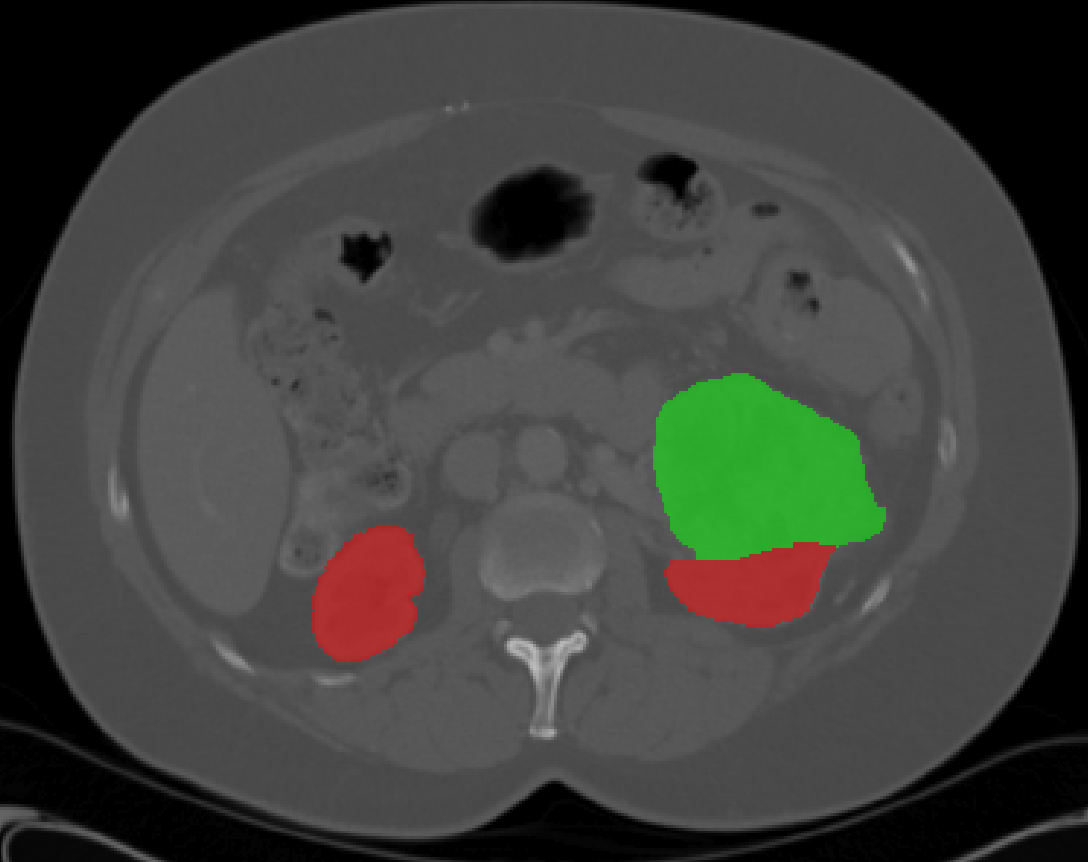}
\hfill
\caption{Example of an axial slice of 3D CT images of two patients in KiTS dataset. 
Red color indicates kidneys, green color indicates tumor region. }
\label{fig:dataexample}
\end{figure*}

\section{Related Work}

Traditionally, various techniques such as deformable models, GrabCuts, region growing and atlas-based methods have been applied to the problem of kidney segmentation. In recent years, researchers have attempted to leverage the power of deep learning and CNNs to build segmentation frameworks that are more automated and less dependant on incorporation of prior shape statistics. Thong et al. \cite{thong2018convolutional} proposed a 2D patch-based approach for kidney segmentation in contrast-enhanced CT scans by leveraging a modified ConvNet. Jackson et al. \cite{jackson2018deep} developed a framework for detection and segmentation and of kidneys in non-contrast CT images by utilizing a 3D U-Net. Yang et al. \cite{yang2018automatic} proposed a method for  kidney and renal tumor segmentation in CT angiography image by a modified residual FCN that is equipped with a pyramid pooling module. Furthermore, Yin et al. \cite{yin2019deep} employed a cascaded approach for segmentation of kidneys with renal cell carcinoma by training a CNN that predicts a bounding box around kidney and a subsequent CNN that segments the kidneys. Recently, Xia et al. \cite{xia2019deep} proposed a two-stage approach for segmentation of kidney and space-occupying lesion area by using SCNN and ResNet for image retrieval and SIFT-flow and MRF for smoothing and pixel matching.   

\label{sec:relatedwork}

\section{Methods}
\label{sec:methods}

\begin{figure*}[t]
  \includegraphics[width=\textwidth]{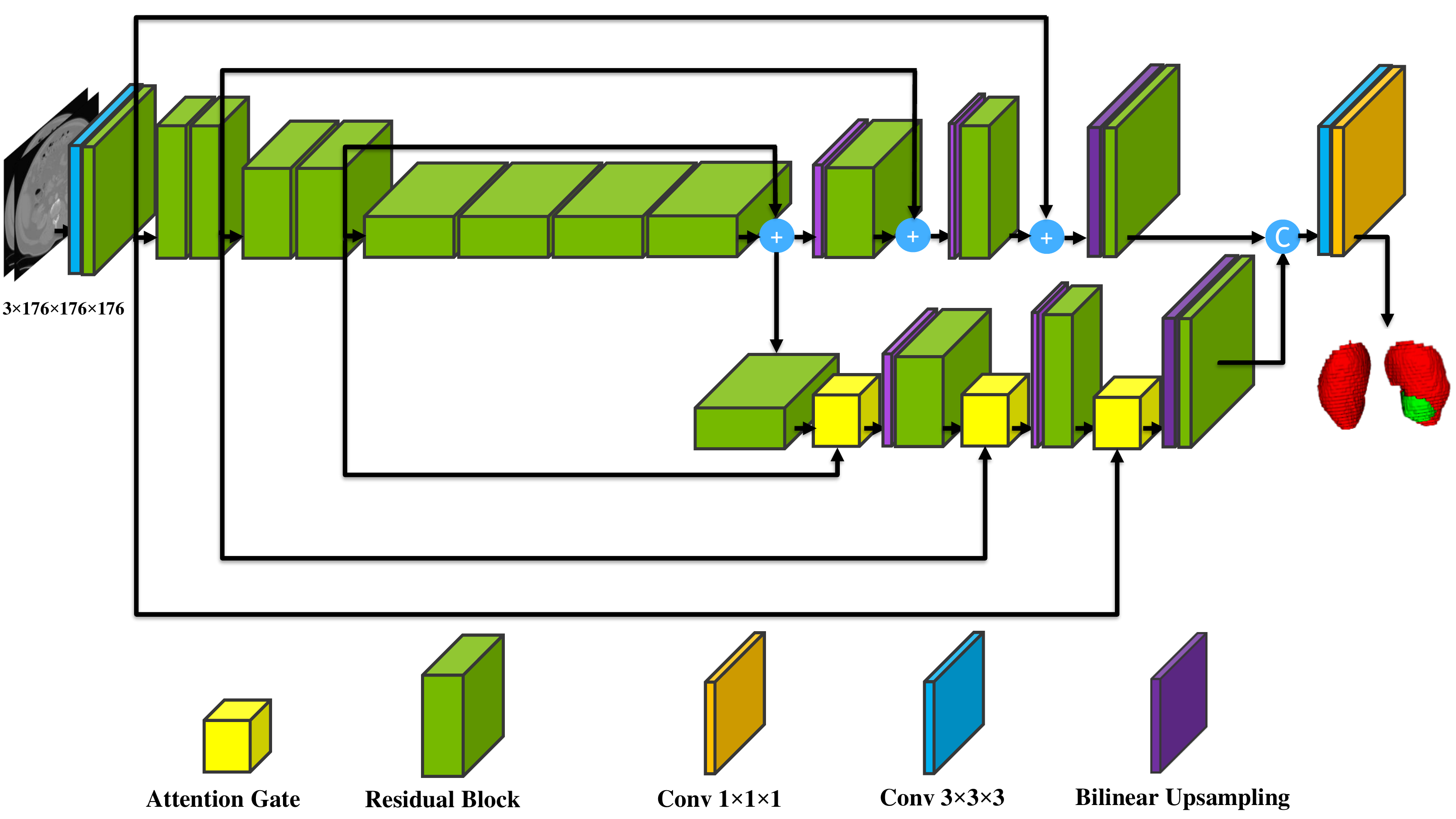}
  \caption{Our proposed CNN architecture.}
  \label{fig:pipeline}
\end{figure*}

\subsection{Framework Architecture}
As illustrated in Figure~\ref{fig:pipeline}, our network consists of the main segmentation branch and the additional boundary stream that processes the feature maps at the boundary level~\cite{hatamizadeh2019boundary}. The main branch follows~\cite{Myronenko18} an asymmetric encoder-decoder structure. The input to the encoder is a 176x176x176 crop which is initially fed into a 3x3x3 convolution with 16 filters. Feature maps are then extracted at each resolution by feeding them into a residual block~\cite{He16} followed by a strided 3x3x3 convolution (for downsizing and doubling of feature dimension). The bottom of the encoder entails four consecutive residual blocks that are connected to the decoder. The extracted feature maps in the decoder are upsampled using bilinear interpolation and added with feature maps from the encoder. The output of the decoder is concatenated with the output of the boundary and fed into a 1x1x1 convolution with 2 channels where channel-wise sigmoid activation $\sigma(X) =\frac{1}{1+\exp({-X})}$ determines the probability of each voxel belonging to kidneys and tumor or only tumor classes.

\subsection{Boundary Stream}
The purpose of the boundary stream is to highlight the edge information of the feature maps extracted in the main encoder by leveraging an additional attention-driven decoder. The attention gates in every resolution of the boundary stream process the feature maps that are learned in the main encoder as well as the output of the previous attention gates. For the first attention gate, we first concatenate the output of the encoder with its previous resolution and feed it into a residual block. In the attention gates, each input is first fed into a 3x3x3 convolutional layer with matching number of feature maps and then fused together, followed by ReLU. The output of the ReLU is fed into a 1x1x1 convolution layer followed by sigmoid function $\sigma$ to obtain the attention map. Consecutively, an element-wise multiplication between the boundary stream feature maps and the computed attention map results in the output of the attention gates.

\subsection{Loss Functions}

We use a dice loss function on the predicted outputs of the main stream as well as the boundary stream. The dice loss is as follows~\cite{Milletari16}:

\begin{equation}
L_{Dice}= 1- \frac{2*\sum y_{true} * y_{pred} }{\sum y_{true}^2 + \sum y_{pred}^2 + \epsilon} 
\label{eq:dice}
\end{equation}   

Where $y_{pred},y_{true}$ denote the voxel-wise semantic predictions of the main stream and their corresponding labels, $\epsilon$ is a small constant to avoid division by zero and summation is carried over the total number of voxels.

Additionally, we add a weighted Binary Cross Entropy (BCE) loss to the boundary stream loss in order to deal with the imbalanced number of boundary and non-boundary voxels: 

\begin{equation}
\begin{split}
&  L_{BCE}= -\beta \sum_{j\in y_{+}} logP(y_{pred,j}=1|x;\theta)\\
& -(1-\beta) \sum_{j\in y_{-}} logP(y_{pred,j}=0|x;\theta)  \\ 
\end{split}
\label{eq:finalloss}
\end{equation}

Where $x,\theta, y_{-}$ and $y_{+}$ denote the 3D input image, CNN parameters, edge and non-edge voxel sets respectively. $\beta$ is the ratio of non-edge pixels over the entire number of voxels and $P(y_{pred,j})$ denotes the probability of the predicated class at voxel $j$. 

The total loss function that is minimized during training is computed by taking the average of losses for tumor-only and foreground class predictions.

\section{Implementation Details and Dataset}

\paragraph{KiTS 2019 dataset}:  Kidney Tumor Segmentation Challenge (KiTS 2019) provides data of multi-phase 3D CTs, voxel-wise ground truth labels, and comprehensive clinical outcomes for 300 patients who underwent nephrectomy for kidney tumors between 2010 to 2018 at University of Minnesota \cite{heller2019kits19}. 210 patients were randomly selected for the training set and the remaining 90 patients were left as a testing set. The annotation was performed in the transverse plane with regular subsampling of series in the longitudinal direction with roughly 50 annotated slices depicting the Kidney for each patient. The labels for excluded slices were computed by using a contour interpolation algorithm~\cite{heller2019kits19}. Figure~\ref{fig:dataexample} illustrates 2D axial view of the example images from two patients in the training set of KiTS 2019.

\paragraph{Data processing}:  We normalized the CT data to [-1, 1] range by dividing the intensity values by $1000$ and clipping the values that fall outside this range. For training, images were re-sampled to 1x1x1mm isotropic resolution and  re-sampled back to their original resolution after the inference. 
The re-sampled output size of the images was on average 512x512 in axial plane and $400-800$ along the inf-sup direction.

\paragraph{Implementation details}:We have implemented our method in Pytorch\footnote{\href{http://pytorch.org/}{http://pytorch.org/}}. Since the re-sampled CT image were often large, we used a 176x176x176 crop during training. The cropping region was centered on the kidney tumor label (with probability 0.8), on any foreground (with probability 0.1) and on background (with probability 0.1). We found it important to sample more frequently from the tumor region.  The model was trained on 8 NVIDIA Tesla V100 16GB GPUs (DGX-1 server). We used a batch size of 8 and the Adam optimization algorithm with the initial learning rate of $ \alpha_{0} = 5e-5$ that was further decreased according to $\alpha = \alpha_{0} *\left(1-e/N_{e}\right)^{0.9}$~\cite{Myronenko18}, where $e$ and $N_{e}$ denote the current epoch counter total number of epochs (300 in our case). During inference, we have leveraged 
test time augmentation (TTA) and have used an ensemble of 5 models to further improve the results. 

\paragraph{Evaluation metrics}: We have adopted the same three evaluation metrics as outlined by KiTS 2019 challenge. Kidneys dice denote the segmentation performance when considering both kidneys and tumors as the foreground whereas tumor dice considers everything except the tumor as background. Composite dice is simply the average of kidneys dice and tumor dice.  


\section{Results and Discussion}

\begin{figure*}[t!]

\includegraphics[width=0.49\linewidth,height=0.49\linewidth]{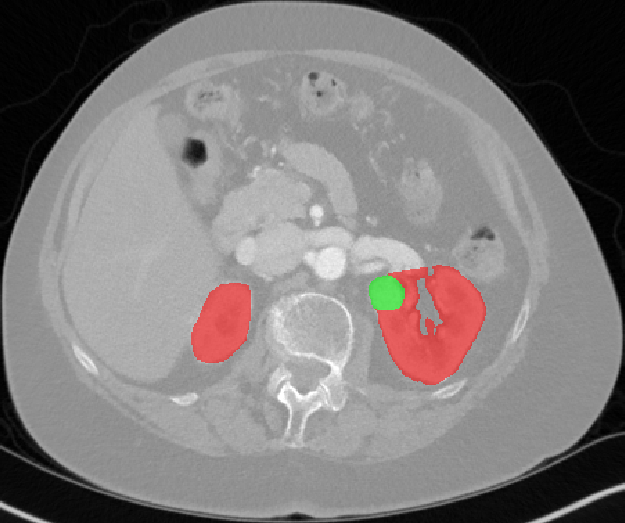}
\hfill
\includegraphics[width=0.49\linewidth,height=0.49\linewidth]{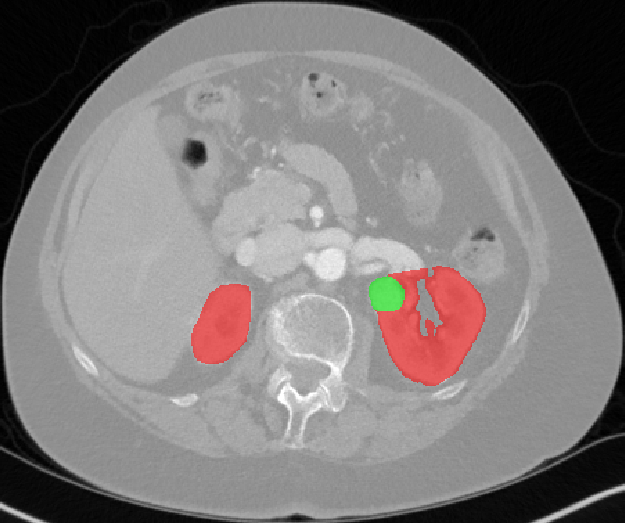}
\hfill

\vspace{1pt}

\includegraphics[width=0.49\linewidth,height=0.49\linewidth]{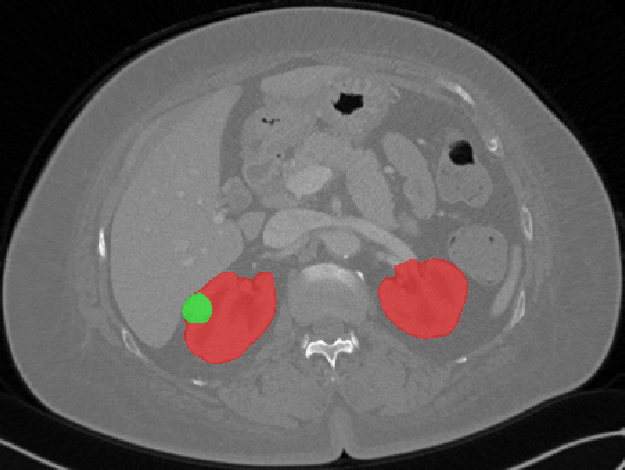}
\hfill
\includegraphics[width=0.49\linewidth,height=0.49\linewidth]{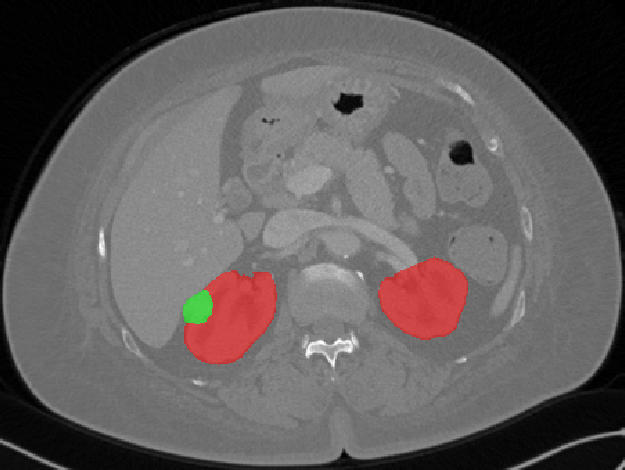}
\hfill

\makebox[0.49\linewidth]{(a) Our Predictions} \hfill \makebox[0.49\linewidth]{(b) Ground truth Labels}

\caption{Visualization of (a) our model's predictions (b) ground truth labels}
\label{fig:res_vis}

\end{figure*}

\paragraph{Preliminary}: Table~\ref{tab:results} represents the evaluation results of our model on our own dataset split. We divided the training set of KiTS 2019 dataset into our own subsets for training and validation and evaluated the performance of a single model as well as an ensemble of 5 models. Finally the KiTS 2019 submission provided two approximate scores based on a small subset of its validation dataset and this allowed us to list the approximate scores of the single model and the ensemble model. Evidently, the dice scores from evaluations on our own split were similar and consistent with the approximate scores provided by the submission portal(see Table~\ref{tab:results}). Figure~\ref{fig:res_vis} illustrates the segmentation visualizations of our method and their corresponding ground truth from two cases in the validation set of our own split.

\begin{table}[t]
	\centering
	\caption{Preliminary dice results based on our own data split as well as 2 approximate scores provided by KiTS 2019 submission portal.}
	\label{tab:results}
	\begin{tabular}{|l|c|c|c|}
		\hline
		Model & Kidneys Dice & Tumor Dice & Composite Dice \\ \hline
		Our split (single model) & 0.957 & 0.821 & 0.889 \\
		Our split (TTA + ensemble) & \textbf{0.970} & \textbf{0.834} & \textbf{0.902} \\
		Approximate score (single model) & 0.955 & 0.736 & 0.845 \\
		Approximate score (TTA + ensemble) & \textbf{0.974} & \textbf{0.784} & \textbf{0.879} \\

		\hline
	\end{tabular}
\end{table}

\paragraph{KiTS 2019 Test Set}: The evaluation results\footnote{\href{http://results.kits-challenge.org/miccai2019/}{http://results.kits-challenge.org/miccai2019/}} of our model on the testing set of KiTS 2019 dataset is presented in Table~\ref{tab:testresults}. The kidneys dice is very similar to approximate scores obtained by the ensemble model that utilizes TTA while the tumor dice is $\%3.35$ better than its counterpart. Moreover, our method ranks 9th overall in terms of the composite dice of kidneys and Tumor among 100 participants in KiTS 2019 challenge. Our model in particular performed better on kidneys segmentation task.

\begin{table}
	\centering
	  \caption{The Evaluation results of our model on KiTS 2019 test set.}
	\label{tab:testresults}
	\begin{tabular}{|c|c|c|}
		\hline
		Kidneys Dice & Tumor Dice & Composite Dice \\ \hline
		0.9742 & 0.8103 & 0.8923 \\
		\hline
	\end{tabular}
\end{table}

\label{sec:conclusion}
\section{Conclusion}
\label{sec:conclusion}

In this work, we have proposed an end-to-end 3D framework for reliable and automated segmentation of kidneys and kidney tumors. Our network consists of a an encoder-decoder architecture equipped with a boundary stream that processes the edge information separately and is supervised by edge-aware losses. We have validated the effectiveness of our approach by training and testing our model on 2019 MICCAI KiTS Kidney Tumor Segmentation Challenge dataset. Our method has achieved dice scores of $0.9742$ and $0.8103$ for kidney and tumor repetitively and an overall composite dice score of $0.8923$ and ranks 9th overall in terms of composite dice among 100 participants of this challenge.

\bibliographystyle{splncs04}
\bibliography{bibliography}

\end{document}